\documentclass[11pt,a4paper]{article}
\usepackage[utf8]{inputenc}
\usepackage[T1]{fontenc}
\usepackage{amsmath}
\usepackage{amsfonts}
\usepackage{amssymb}
\usepackage{graphicx}
\usepackage[round]{natbib}
\usepackage{booktabs}
\usepackage[margin=2.5cm]{geometry}

\begin{document}
	\title{Functional and variables selection in extreme value models for regional flood frequency analysis}
	\author{Aldo Gardini\thanks{Department of Statistical Sciences, University of Bologna. E-mail: \texttt{aldo.gardini@unibo.it}}}
	\date{}
	\maketitle
	\begin{abstract}
		\noindent The problem of estimating return levels of river discharge, relevant in flood frequency analysis, is tackled by relying on the extreme value theory. The Generalized Extreme Value (GEV) distribution is assumed to model annual maxima values of river discharge registered at multiple gauging stations belonging to the same river basin. The specific features of the data from the Upper Danube basin drive the definition of the proposed statistical model. Firstly, Bayesian P-splines are considered to account for the non-linear effects of station-specific covariates on the GEV parameters. Secondly, the problem of functional and variable selection is addressed by imposing a grouped horseshoe prior on the coefficients, to encourage the shrinkage of non-relevant components to zero. A cross-validation study is organized to compare the proposed modeling solution to other models, showing its potential in reducing the uncertainty of the ungauged predictions without affecting their calibration.
	\end{abstract}
	\textbf{Keywords:} Bayesian P-splines, Generalized extreme value distribution, Horseshoe prior, Return levels, Stan

	\section{Introduction}
	
	An effective prediction of flood phenomena is crucial for protecting and managing territories. A rigorous hydrological approach to the problem can be profitably supported by flood frequency analysis, which is a data-based framework aimed at providing reliable estimates of expected return periods of a flood event characterized by a certain magnitude. When data from multiple gauging stations placed in a target catchment or area are included in the analysis, a regional flood frequency analysis is carried out \citep{hosking_wallis_1997}. The main advantage of this approach is the possibility to borrow strength from available stations to obtain calibrated and reliable estimates also for ungauged locations. 
	In the framework of regional flood frequency analysis, a pioneering approach is the flood index by \citet{dalrymple1960flood}. Such a method is characterized by a multi-step procedure, where the main stages can be summarised in \textit{i)} classifying stations in homogeneous regions; \textit{ii)} choosing a suitable frequency distribution for the locations included in a region; \textit{iii)} estimating the distribution parameters, commonly relying on $L$-moments algorithm. For an overview of this approach, see \citet{hosking_wallis_1997}. 
	
	An alternative strategy that allows producing return level estimates also for ungauged locations relies on modeling a block maxima sequence through a Generalized Extreme Value (GEV) distribution, assuming that the parameters depend on covariates and/or other features connected with the stations. In flood frequency analysis, yearly maxima of the river discharge ($m^3/s$) are considered for this purpose. The GEV distribution is a widespread tool of the extreme value theory, and such models have been already exploited in regional flood frequency analysis, often adopting a Bayesian inferential approach. For example, \citet{thorarinsdottir2018bayesian} use it to study floods in Norway, including some features of the stations and the related sub-catchments. A similar modeling strategy is also set by \citet{johannesson2022approximate}, which propose a computationally efficient procedure for its estimation, exploiting its representation as a generalized latent Gaussian model. They also include in the predictors spatially structured random effects, as done, among the others, by \citet{dyrrdal2015bayesian} and \citet{geirsson2015computationally} in modeling precipitation extremes. This model architecture is also used, for example, by \citet{huerta2007time} to analyze Ozone concentration extremes, \citet{lee2013bayesian} to model wind data and \citet{raty2022bayesian} for sea levels. A limitation of such a strategy is caused by the conditional independence assumption among stations: this allows only to predict marginal return levels. If multivariate return levels are needed, a max-stable modeling framework should be pursued \citep{asadi2015extremes}.
	
	In flood frequency analysis, this class of GEV regression models generally assumes linearity among the covariates and parameters.  
	This could represent an important restriction in the analysis of complex environmental processes that can be relaxed by defining flexible models based on spline regression. In this paper, the setting of Bayesian P-splines by \citet{lang2004bayesian} is adopted. Its convenience is due to the parsimonious parameterization brought by the usage of basis functions and the automatic penalization for roughness induced by the use of smoothing priors for splines coefficients \citep{fahrmeir2010bayesian}. \citet{raty2022bayesian} proposed their usage in modeling sea levels extremes.  Other examples of extreme values models that include spline regressions can be found in \citet{lee2013bayesian}, which exploited Bayesian multivariate adaptive regression spline in modeling extreme loads in wind turbines and \citet{yousfi2017regularized} which discuss and compare different penalization methods for B-splines. Lastly, it is worth mentioning the body of literature focusing on frequentist spline models, such as \citet{chavez2005generalized}, \citet{padoan2008mixed} and \citet{rohmer2021revisiting}, for which interesting computational tools are also provided \citep[e.g., the \texttt{evgam} package][]{youngman2022evgam}.
	
	In this paper, data from stations located in the Upper Danube River basin are analyzed with the aim of carrying out a regional flood frequency analysis. The exploratory analysis pointed out that the relationship between station-specific covariates and the GEV parameters is strongly non-linear, motivating the proposal of a GEV regression model with Bayesian P-splines. In this framework, another interesting problem is the selection of relevant regressors. For example, \citet{dyrrdal2015bayesian} carry out this step through a Bayesian model averaging component in a regional flood frequency analysis model that assumes linearity. The use of P-splines poses an additional problem of function selection, in order to obtain a model with only relevant covariates showing a parsimonious representation of their effect, i.e., their impact on parameters. This task was tackled by \citet{scheipl2012spike}, which proposed a particular formulation of spike-and-slab prior that hierarchically performs both the selection steps at once. An interesting prior distribution that is able to mimic the behavior of the spike-and-slab prior is the horseshoe \citep[HS,][]{carvalho2010horseshoe}, which does not introduce discrete latent variables and, for this reason, it is also implementable within the popular \texttt{Stan} probabilistic language \citep{carpenter2017stan}. The HS prior can be extended to define a grouped HS prior \citep{xu2016bayesian} that is able to perform both variable and functional selection. Such a prior distribution is adopted for the coefficients involved in the GEV regression with P-splines and its effectiveness in improving the predictions of return levels for ungauged locations is discussed by means of a cross-validation study.
	
	The rest of the paper is organized as follows. Section~\ref{sec:GEV} contains an introduction to extreme value theory and the GEV distribution, setting also basic notations. The Danube data are introduced in Section~\ref{sec:data}, together with an exploratory analysis that motivates the development of the proposed modeling solutions, which are defined in Section~\ref{sec:model}. The empirical results coming from a cross-validation study and from the analysis of the whole dataset are shown in Section~\ref{sec:empirical}, whereas Section~\ref{sec:concl} offers some concluding remarks.

	\section{Basic concepts of extreme value theory}\label{sec:GEV}
	When the main interest of a statistical procedure is to describe a phenomenon through quantities strongly related to the tails of the distribution, then it is necessary to resort to the extreme value theory. In this branch of statistics, two main approaches can be pursued: block maxima and peak-over-threshold \citep{coles2001introduction,beirlant2004statistics}. The first strategy considers the maxima of a time block sequence, which are used to estimate the parameters of the GEV distribution. The second procedure is constituted by two steps: \textit{i)} the whole dataset is exploited to estimate a threshold above which observations are considered to be extremes; \textit{ii)} the threshold exceedances are used to fit a Generalized Pareto distribution, that it can be shown to be connected with the GEV.
	
	In this paper, the block maxima approach is adopted. According to the Fisher - Tippett – Gnedenko theorem, the GEV distribution arises as the limiting distribution of a normalized sequence of block maxima, and, hence, plays a relevant role in this framework. Indeed, this probabilistic result is exploited in extreme value theory, assuming that a sequence of recorded maxima over $T$ distinct temporal blocks (e.g. years or days), denoted by $y_t,\ t=1,\dots,T$, is distributed as
	\begin{equation}\label{eq:dist}
		y_t|\mu,\sigma,\xi\stackrel{ind}{\sim}GEV(\mu,\sigma,\xi),\quad \forall t.   
	\end{equation}
	Such a distribution is ruled by three parameters: $\mu\in\mathbb{R}$ controls the location, $\sigma\in\mathbb{R}^+$ the scale and $\xi\in\mathbb{R}$ the shape, affecting the behavior of the distribution tails and, consequently, its support. In particular, $\xi<0$ implies a short and finite right-tail ($y_t\in(-\infty;\mu-\sigma/\xi]$), $\xi=0$ a light right-tail ($y_t\in\mathbb{R}$) and $\xi>0$ an heavy right-tail ($y_t\in[\mu-\sigma/\xi,+\infty)$). In the latter case, $\xi$ has an impact also on the existence of the distribution moments: the moment of order $\rho$ is finite if $\xi<1/\rho$. 
	The GEV distribution is usually defined through its cumulative distribution function:
	\begin{equation}\label{eq:cdf}
		F(y;\mu,\sigma,\xi)=\left\{\begin{array}{ll}
			\exp\left\{-\left[1+\xi\left(\frac{y-\mu}{\sigma}\right)\right]_+^{-\frac{1}{\xi}}\right\},  & \xi\neq 0; \\
			\exp\left\{-\exp\left\{-\frac{y-\mu}{\sigma}\right\} \right\}, & \xi=0;
		\end{array} \right.
	\end{equation}
	where $[g]_+=\max(g,0)$. 
	
	In the statistical analysis of extremes, the most typical output is the estimation of return levels associated with a return period $R$. It is defined as the quantile $Q_p$ that has a probability equal to $p=1/R$ of being exceeded in the chosen time block. In other words, the return level $Q_{1/R}$ is expected to be surpassed once every $R$ time blocks and, inverting the \eqref{eq:cdf}, it is defined as
	\begin{equation}\label{eq:rp}
		Q_{1/R}=\left\{\begin{array}{ll}
			\mu-\frac{\sigma}{\xi}\left[1+\log(1-1/R)^{-\xi}\right],  & \xi\neq 0; \\
			\mu-\sigma\log\left[-\log(1-1/R)\right], & \xi=0.
		\end{array} \right.
	\end{equation}
	
	Once the probabilistic setting is defined, some remarks about the inferential side of the problem are worthy. In this paper, a Bayesian approach is adopted: it is becoming increasingly popular in extremes statistics thanks to the possibility of eliciting prior information and the natural ability to estimate the model uncertainty, propagating it in distinct steps of the analysis \citep{coles1996bayesian,coles2001introduction}. For example, making inference on return levels \eqref{eq:rp} require combining the three distribution parameters that need to be estimated: if the Bayesian approach is chosen and Monte Carlo Markov Chain (MCMC) methods are exploited, then drowns from the parameters posterior can be combined to obtain the whole posterior distribution of $Q_{1/R}$. 
	
	\section{Data on Danube river basin}\label{sec:data}
	
	\begin{figure}
		\centering
		\includegraphics[width = .9\linewidth]{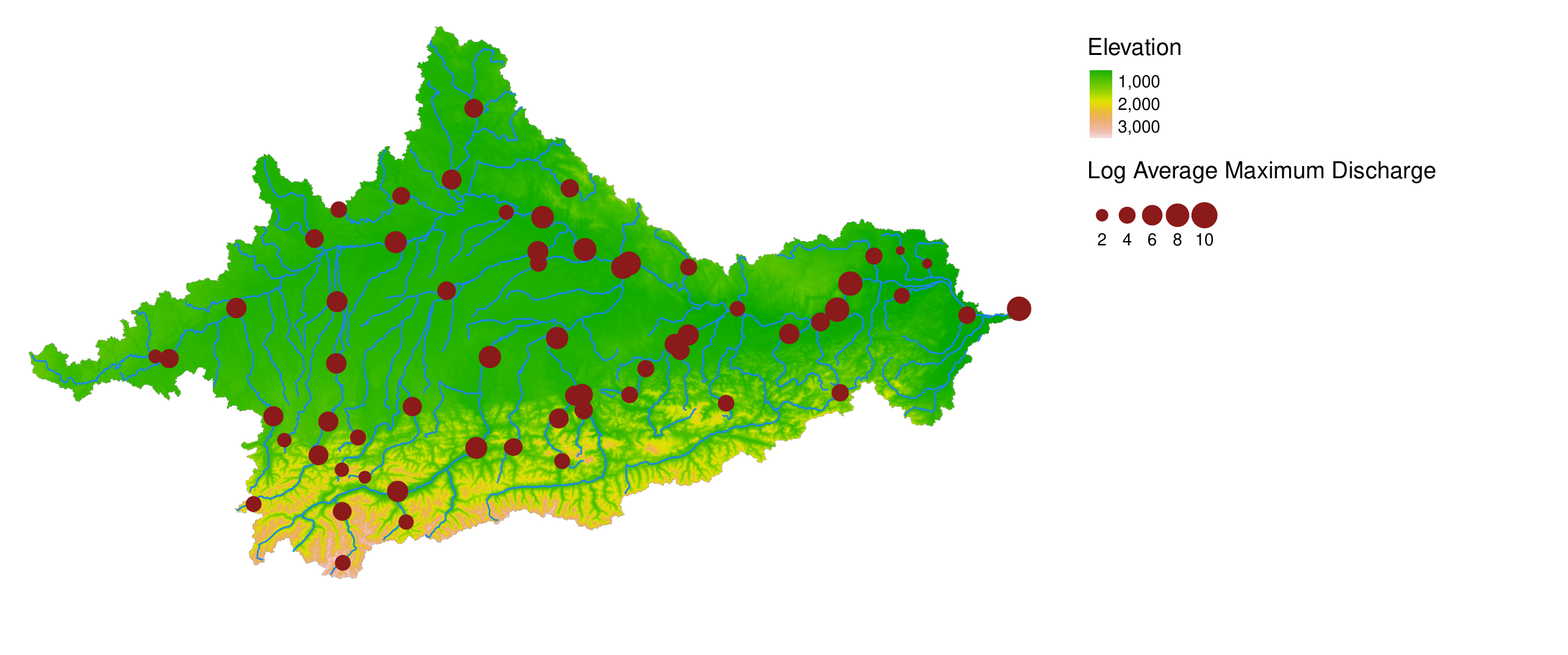}
		\caption{Gauging stations of the Danube upper basin included in the analysis.}
		\label{fig:area}
	\end{figure}

	The proposed strategy targets the estimation of return levels for the discharge of rivers belonging to the Danube upper basin (i.e. the part located both in Germany and Austria). The analysis considers data that are freely available from different sources to propose a general procedure that can also be replicated in other river basins. In this section, the sources of information considered in the analysis are listed, together with some remarks about the data integration procedure.
	
	\begin{table}[]
		\centering
		\begin{tabular}{@{}llll@{}}
			\toprule
			Covariate & Description                       & Source & Transformation  \\ \midrule
			Latitude &  Station latitude. &  GRDC  &  Identity\\
			Longitude &  Station longitude. &  GRDC  &  Identity\\
			Area      & Area of the station sub-catchment. & EU-DEM & Logarithm  \\
			Elevation  & Mean terrain elevation.  &  EU-DEM &  Logarithm  \\
			Slope    &  Mean slope.   &  EU-DEM  & Logarithm  \\
			Aspect    &  Mean aspect of slopes.   &  EU-DEM  & Identity  \\
			Cover    &  Proportion of built areas.   &  CORINE  & Identity  \\
			Rainfall    &  Mean annual rainfall.   &  WorldClim  & Logarithm  \\\bottomrule
		\end{tabular}
		\caption{Station-specific covariates used in the analysis.}
		\label{tab:cov}
	\end{table}
	
	The response variable required for implementing a flood frequency analysis is the river discharge, usually measured in $m^3/s$. The time series with daily river discharge observations are retrieved from the GRDC portal \citep{grdc}, selecting all the gauging stations present in the area under study. To determine the final set of locations, some data quality and reliability checks are performed: by focusing on the period 1985-2017, only stations with a maximum of 2\% daily missing observations in each year are selected. Furthermore, the coordinates of the gauges should correctly locate on the river network to avoid location inconsistencies and possible mismatches. The shapefiles with the network of main rivers in the basin are retrieved from the River Network Database \citep{rivers}. The final database is constituted by yearly maxima from $S=62$ stations that satisfied the aforementioned requirements. The yearly maxima are denoted as $y_{st}$, referring to the gauging station $s=1,\dots,62$ in the year $t=1,\dots,33$. For brevity, $\mathbf{y}_s$ indicates the vector of maxima related to a single station $s$. 
	
	River discharge values are complemented by station-specific auxiliary variables, that are listed in Table~\ref{tab:cov}. Most of them are derived from the EU Digital Elevation Model \citep[EU-DEM,][]{dem}, starting from the determination of stations sub-catchments through GIS-based tools within the \texttt{whitebox} \texttt{R} package \citep{lindsay2016whitebox}. The spatial location is accounted for by latitude and longitude, the sub-catchment area is an important measure directly related to the river discharge magnitude, whereas the features of the terrain characterizing the sub-catchment are computed by averaging indicators derived from the EU-DEM in the area (elevation, slope and aspect). In addition, the average rainfall \citep{o2012bioclimatic} and the proportion of area covered by buildings \citep[from CORINE land cover raster,][]{buchhorn2020copernicus} are considered.

	\subsection{Exploratory analysis}\label{sec:expl}
	
	This section reports the results of an exploratory analysis aimed at pointing out the main motivations for the modeling strategies dealt with in the paper.  A two-step analysis is performed: firstly, a GEV distribution is fitted on the maxima sequences $\mathbf{y}_s$ registered at each station $s$: $y_{st}\stackrel{ind}{\sim}GEV(\mu_s,\sigma_s,\xi_s),\ \forall t$. Then, the estimates of the GEV parameters are used as responses in Bayesian semi-parametric additive models, to investigate how they are influenced by the station-specific covariates.
	
	\begin{figure}
		\centering
		\includegraphics[width = \linewidth]{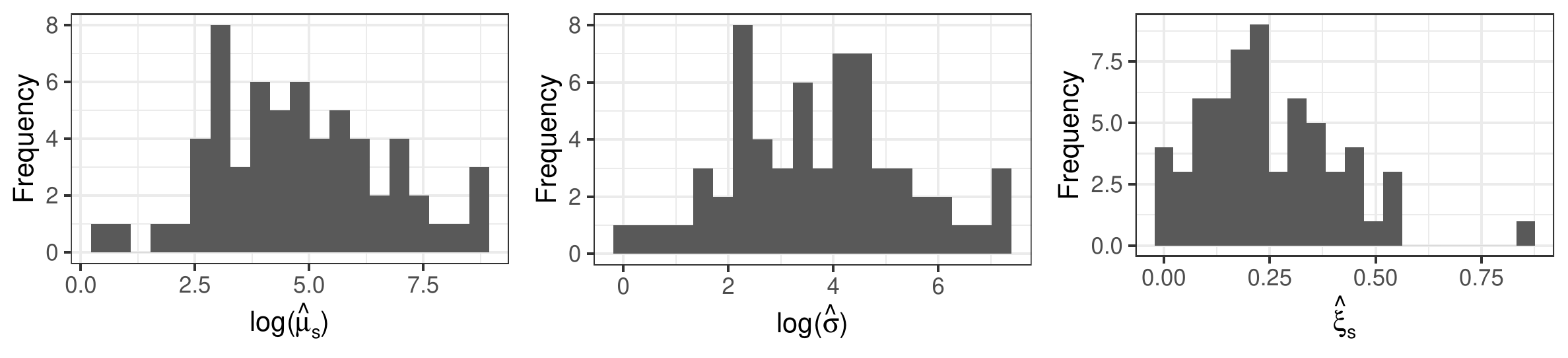}
		\caption{Histograms of posterior means for the station-specific GEV parameters.}
		\label{fig:hist_explo}
	\end{figure}
	
	In the first step, $S=62$ station-specific GEV models are fitted adopting the Bayesian approach. To retrieve the posterior distributions of the parameters, the model specification must be completed by choosing prior distributions. Given the exploratory purpose of this step, non-informative priors on the parameters are set; namely, zero mean Gaussian distributions with scale 10,000 for $\mu_s$ and $\log(\sigma_s)$, and a standard normal for the shape parameter $\xi_s$. Samples from the posterior distribution of the GEV parameters are drawn by means of an MCMC algorithm implemented in \texttt{Stan}. As point estimates, the posterior means are then computed and are denoted as $(\hat\mu_s,\hat\sigma_s,\hat\xi_s),\ \forall s$. 
	
	Figure~\ref{fig:hist_explo} reports the distributions of the estimated GEV parameters, which are contained in the following vectors: $\hat{\boldsymbol\mu}=(\hat\mu_1,\dots,\hat\mu_S)^\top$, $\hat{\boldsymbol\sigma}$, $\hat{\boldsymbol\xi}$. The first goal of this exploratory step is to assess whether the multivariate link function proposed in \citet{johannesson2022approximate} is reasonable for the discussed application. We first note that all the MCMC draws from the posteriors of $\mu_s,\ \forall s$, are positive, and the distribution of $\hat{\boldsymbol\mu}$ is markedly skewed. For this reason, a logarithmic transformation of the location parameter might be useful to reduce the skewness, noting that the induced positivity assumption is not restrictive. Furthermore, Pearson's linear correlation between $\log(\hat{\boldsymbol\mu})$ and $\log(\hat{\boldsymbol\sigma})$ is 0.987: consequently, modeling the functional $\tau_s=\log\left(\sigma_s/\mu_s\right)$ as dispersion parameter in place of $\sigma_s$ is convenient to reduce the dependency between model parameters. The multivariate link function by \citet{johannesson2022approximate} also encloses a transformation of the shape parameter $h(\xi_s)$, by limiting its range in the interval $(-0.5,0.5)$. Such a transformation, which will be deeply discussed in the next section, is helpful in stabilizing the estimate of the shape parameter and it appears to be tenable as only 4 stations show an estimate $\hat{\xi}_s$ higher than 0.5.
	
	\begin{figure}
		\centering
		\includegraphics[width = \linewidth]{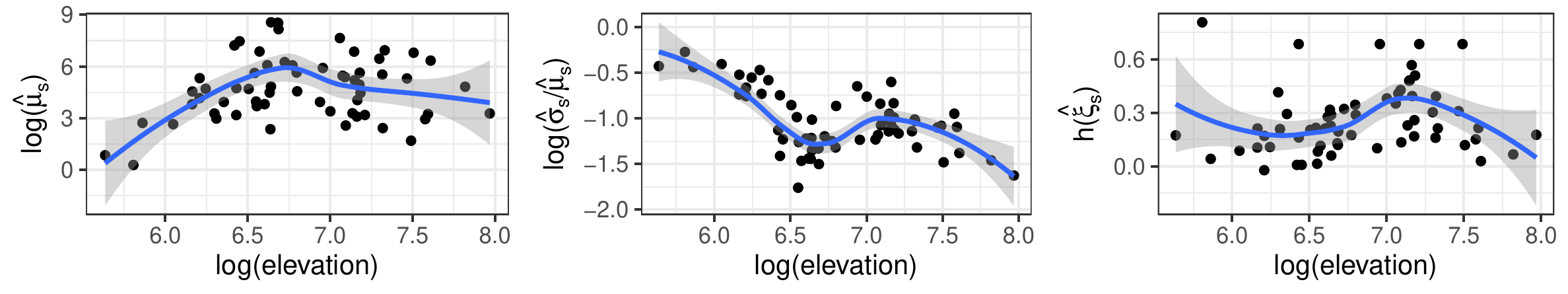}
		\caption{Effect of the logarithm of the elevation on the transformed parameters estimates $\log(\hat{\boldsymbol\mu})$, $\log(\hat{\boldsymbol\sigma}/\hat{\boldsymbol\mu})$ and $h(\hat{\boldsymbol\xi})$.}
		\label{fig:post_expl}
	\end{figure}
	
\begin{figure}
		\centering
		\includegraphics[width = \linewidth]{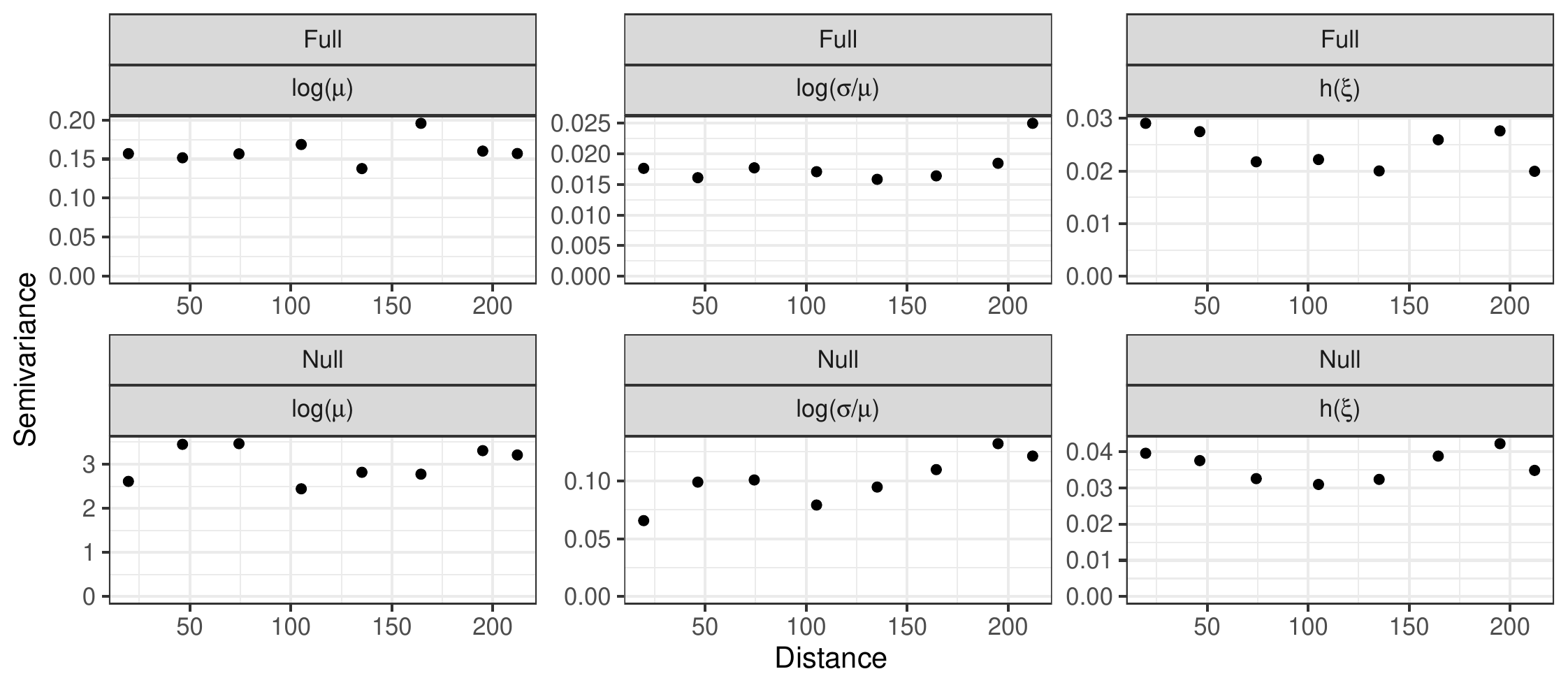}
		\caption{Variograms for the residuals of the models fitted with (\textit{Full}) or without (\textit{Null}) covariates on the transformations of the GEV parameters.}
		\label{fig:variog}
	\end{figure}

	In the second step of the analysis, the transformations of GEV parameters estimates, i.e. $\log(\hat{\boldsymbol\mu})$, $\log(\hat{\boldsymbol\sigma}/\hat{\boldsymbol\mu})$ and $h(\hat{\boldsymbol\xi})$, are used as responses in three distinct Bayesian Gaussian additive models implemented by the \texttt{stan\_gamm4()} function of the \texttt{rstanarm} package \citep{rstanarm}. All the covariates summarized in Table~\ref{tab:cov}, transformed according to the reported function, are included in the model as smooth terms through a Bayesian P-spline representation. This choice is motivated by the presence of non-linear relationships among the covariates and the target parameters (e.g., see Figure~\ref{fig:post_expl}). The exploration of the posterior results supports the choice of specifying smooth effects for the covariates. 
	The residuals of the fitted additive models are studied in order to check if a residual spatial trend can be detected \citep{cooley2007bayesian}. The variograms reported in Figure~\ref{fig:variog} are related both to the model with all the covariates (\textit{Full}) and the model without covariates (\textit{Null}) and they do not point out a relevant residual spatial variation when the full model is considered.
	
	For these reasons, our modeling proposal focuses mainly on the presence of non-linear relationships among the transformations of the GEV parameters and the covariates included in the analysis. On the other hand, the inclusion of spatially structured random effects is omitted to keep the model as simple as possible, pointing the attention on the implementation of Bayesian semi-parametric GEV models.

	\section{The proposed modeling framework}\label{sec:model}
	
	Let us consider that a collection of $N$ maxima $y_{st}$ from $s=1,\dots,S$ gauging stations are available for blocking times $t=1,\dots,T$. It is assumed that, conditionally on site-specific parameters, the maxima are distributed as:
	\begin{equation}
		y_{st}|\mu_s,\sigma_s,\xi_s\stackrel{ind}{\sim}GEV(\mu_s,\sigma_s,\xi_s),\ \forall s, t.
	\end{equation}
	The assumption of conditional independence represents a simplification but it is a quite standard one in the extreme value literature when marginal return levels need to be estimated \citep{dyrrdal2015bayesian,thorarinsdottir2018bayesian,johannesson2022approximate}. Alternatively, max-stable spatial processes can be considered, even if the complexity of the modeling framework sensibly increases \citep{asadi2015extremes}.
	As already hinted in Section~\ref{sec:expl}, the multivariate link function proposed in \citet{johannesson2022approximate} is adopted, to obtain transformed parameters that are convenient to specify regression models for:
	\begin{equation*}
		g(\mu_s,\sigma_s,\xi_s)=\left(\psi_s = \log(\mu_s), \tau_s = \log(\sigma_s/\mu_s), \phi_s = h(\xi_s)\right)^\top.
	\end{equation*}
	The station-specific parameters are stored in the following vectors: $\boldsymbol\psi = (\psi_1,\dots,\psi_S)^\top$, $\boldsymbol\tau = (\tau_1,\dots,\tau_S)^\top$ and $\boldsymbol\phi = (\phi_1,\dots,\phi_S)^\top$. Applying the logarithmic transformation on the location parameter $\mu_s$ restricts its domain to $\mathbb{R}^+$ and, according to the exploratory results of Section \ref{sec:expl}, such an assumption is fulfilled in the discussed application; furthermore, the dependence between scale and location parameters is mitigated modeling $\log(\sigma_s/\mu_s)$ instead of $\log(\sigma_s)$. Lastly, the function applied to the shape parameter is
	\begin{equation}
		h(\xi_s)=a_\phi+b_\phi\log\left\{-\log\left[1-(\xi+0.5)^{c_\phi}\right]\right\},
	\end{equation}
	where $(a_\phi,b_\phi,c_\phi)=(0.062376,0.39563,0.8)$. \citet{johannesson2022approximate} proposed it to restrict the domain of $\xi_s$ to the interval $(0.5,0.5)$, keeping an approximately linear relationship with $\xi_s$ around zero. The domain restriction for the shape parameter is motivated by the desiderata of guaranteeing the variance of the GEV distribution to be finite and the upper bound of the distribution greater than $\mu_s+2\sigma_s$.
	Note that the developments discussed later still hold even if a different link function is adopted. 
	
	If a linear relationship between covariates and the parameters is assumed, the following latent regression models are specified:
	\begin{equation}\label{eq:lin}
		\begin{aligned}
			&\boldsymbol\psi = \boldsymbol{1}_S\beta_{0\psi}+\mathbf{X}\boldsymbol{\beta}_{\psi}+\mathbf{u}_\psi;\\
			&\boldsymbol\tau = \boldsymbol{1}_S\beta_{0\tau}+\mathbf{X}\boldsymbol{\beta}_{\tau}+\mathbf{u}_\tau;\\
			&\boldsymbol\phi = \boldsymbol{1}_S\beta_{0\phi}+\mathbf{X}\boldsymbol{\beta}_{\phi}+\mathbf{u}_\phi;
		\end{aligned}
	\end{equation}
	where the design matrix $\mathbf{X}=[\mathbf{x}_{\bullet1}\cdots \mathbf{x}_{\bullet M}]\in\mathbb{R}^{S\times M}$ contains the standardized covariates $\mathbf{x}_{\bullet m}\in \mathbb{R}^S,\ m=1,\dots,M$.
	Each predictor, related to the generic parameters vector $\boldsymbol{\theta}\in\{\boldsymbol\psi,\boldsymbol\tau,\boldsymbol\phi\}$, is constituted by an overall intercept $\beta_{0\theta}$ and a linear regression with coefficients stored in $\boldsymbol{\beta}_{\theta}\in\mathbb{R}^{M}$. A vector of station-specific unstructured random effects $\mathbf{u}_\theta\in\mathbb{R}^S$ completes the equation, to account for possible residual variation.
	To keep the notation simple, all the model equations in \eqref{eq:lin} contain the same covariates, as will be the case in the application, but this assumption can be easily relaxed. 
	
	The model specification must be completed by setting prior distributions for the parameters. Firstly, weakly informative Gaussian priors are assumed for the coefficients. To account that the transformed GEV parameters have different magnitudes, they are calibrated by exploiting the results from the exploratory analysis of Section~\ref{sec:expl}. More in detail, following the advises from the \texttt{rstanarm} package \citep{rstanarm}, the following prior is set for the intercepts:
	\begin{equation}\label{eq:priorbeta0}
		\beta_{0\theta}\sim\mathcal{N}\left(m_{\hat{\theta}},2^2s^2_{\hat\theta}\right),
	\end{equation}
	where $m_{\hat{\theta}}$ and $s^2_{\hat\theta}$ are the mean and the variance of the generic vector of fitted parameters $\hat{\boldsymbol{\theta}}$. Recalling that the covariates are standardized, independent zero-mean Gaussian priors with equal scales are specified for the regression coefficients:
	\begin{equation}\label{eq:priorbeta}
		\beta_{\theta m }\sim\mathcal{N}\left(0,2^2\right),\ m=1,\dots,M.
	\end{equation}
	Lastly, focusing on the vector of unstructured random effects, a spherical multivariate Gaussian prior with scale parameter $\kappa_{\theta}$ is set:
	\begin{equation}\label{eq:prioru}
		\mathbf{u}_{\theta}|\kappa_{\theta}\sim\mathcal{N}_S(\boldsymbol{0},\kappa_{\theta}^2\mathbf{I}_S), \quad \kappa_{\theta}\sim \mathcal{N}^+(0,2^2), \quad \theta\in\{\psi,\tau,\phi\};    
	\end{equation}
	where $\mathcal{N}^+(\cdot, \cdot)$ indicates an half-Normal distribution.
	
	\subsection{GEV regression with Bayesian P-splines}
	
	When the evidence of non-linear relationships between covariates and responses is pointed out, it is natural to extend the linear models in \eqref{eq:lin} by allowing for flexible regression terms. Among the possible strategies, the Bayesian P-splines method by \citet{lang2004bayesian} is implemented:
	\begin{equation}\label{eq:m1}
		\begin{aligned}
			&\boldsymbol\psi = \boldsymbol{1}_S\beta_{0\psi}+\sum_{m=1}^M\mathbf{B}_{m}\boldsymbol\gamma_{\psi,m}+\mathbf{u}_\psi;\\
			&\boldsymbol\tau = \boldsymbol{1}_S\beta_{0\tau}+\sum_{m=1}^M\mathbf{B}_{m}\boldsymbol\gamma_{\tau,m}+\mathbf{u}_\tau;\\
			&\boldsymbol\phi = \boldsymbol{1}_S\beta_{0\phi}+\sum_{m=1}^M\mathbf{B}_{m}\boldsymbol\gamma_{\phi,m}+\mathbf{u}_\phi.
		\end{aligned}
	\end{equation}
	The predictors are characterized by the sum of $M$ flexible regression terms defined as the product of a matrix $\mathbf{B}_{m}\in\mathbb{R}^{S\times K}$ of cubic B-spline basis functions evaluated at $K$ knots, multiplied by a vector of associated coefficients $\boldsymbol\gamma_{\theta m}\in\mathbb{R}^K$. In the P-splines approach, the smoothness of the fitted effect is encouraged by setting a second-order random walk prior on the splines coefficients:
	\begin{equation}\label{eq:prior_spl}
		\boldsymbol{\gamma}_{\theta m}|\omega_{\theta m}\sim\mathcal{N}_K(\boldsymbol{0},\omega_{\theta m}^2\mathbf{K}_\gamma^-), \quad \omega_{\theta m}\sim \mathcal{N}^+(0,2^2), \quad \forall m;\ \theta\in\{\psi,\tau,\phi\}.
	\end{equation}
	The matrix $\mathbf{K}_\gamma$ has rank $K-2$ and it is a precision matrix describing a second-order random walk, whereas $\omega_{\theta m}$ is a scaling parameter. Due to the rank deficiency of the precision matrix, the prior is improper and the specification of linear constraints might be required. To better understand the features of the P-splines setting, the representation of the \eqref{eq:m1} as a mixed model could be useful.
	
	\subsubsection{Mixed model representation}
	
	The linear predictors defined in \eqref{eq:m1} can be reparameterized by exploiting the spectral decomposition of $\mathbf{B}_{m}\mathbf{K}_\gamma^-\mathbf{B}_{m}^\top$, i.e. the covariance matrix of $\mathbf{B}_{m}\boldsymbol{\gamma}_{\theta m}$. The model representation defined in the following is particularly suitable to perform functional selection since the structured and improper prior on the spline coefficients in \eqref{eq:prior_spl} is traced back to a proper spherical Gaussian prior on coefficients, associated with a matrix of orthonormal basis \cite{scheipl2012spike}.
	
	To set the notation, the spectral decomposition is defined as:
	\begin{equation}\label{eq:decomp}
		\mathbf{B}_{m}\mathbf{K}_\gamma^-\mathbf{B}_{m}^\top=\begin{bmatrix}
			\mathbf{U}_+&\mathbf{U}_0
		\end{bmatrix}^\top
		\begin{bmatrix}
			\boldsymbol{\Lambda}_+&\boldsymbol{0}\\
			\boldsymbol{0} & \boldsymbol{0} \\
		\end{bmatrix}
		\begin{bmatrix}
			\mathbf{U}_+&\mathbf{U}_0
		\end{bmatrix}=\mathbf{U}_+\boldsymbol{\Lambda}_+\mathbf{U}_+^\top,    
	\end{equation}
	where $\boldsymbol{\Lambda}_+\in \mathbb{R}^{(K-2)\times(K-2)}$ is a diagonal matrix containing the non-null eigenvalues, $\mathbf{U}_+\in\mathbb{R}^{S\times(K-2)}$ is the orthogonal matrix with the associated eigenvectors and $\mathbf{U}_0\in\mathbb{R}^{S\times(R-K+2)}$ contains the eigenvalues that span the null space of $\mathbf{B}_{m}\mathbf{K}_\gamma^-\mathbf{B}_{m}^\top$.

	Combining the prior in \eqref{eq:prior_spl} and the spectral decomposition \eqref{eq:decomp}, it is possible to split the generic flexible term $\mathbf{B}_m\boldsymbol{\gamma}_{\theta m}$ into a penalized component and an unpenalized one:
	$$
	\mathbf{B}_m\boldsymbol{\gamma}_{\theta m} = \mathbf{x}_{\bullet m}\beta_{\theta m}+\tilde{\mathbf{B}}_{m}\boldsymbol{\tilde{\gamma}}_{\theta m}.
	$$
	The unpenalized part is constituted by the term $\mathbf{x}_{\bullet m}\beta_{\theta m}$ and it is strictly related to the null space of the structure matrix that defines the prior assumed for the splines coefficients. Indeed, under the considered second-order random walk, a polynomial of order one in the covariate is required, i.e. a constant term (already included in the overall intercept) and a linear trend on the covariate. Concerning the penalized component, $\tilde{\mathbf{B}}_{m} = \mathbf{U}_+\boldsymbol{\Lambda}_+^{\frac{1}{2}}\in \mathbb{R}^{S\times(K-2)}$ determines a matrix of orthonormal basis and $\tilde{\boldsymbol{\gamma}}_{\theta m}\in \mathbb{R}^{K-2}$ constitutes the vector of related splines coefficients. Due to the orthogonalization procedure, such vector of coefficients has a spherical prior: $\tilde{\boldsymbol{\gamma}}_{\theta m}|\omega_{\theta m}\sim\mathcal{N}_{K-2}(\boldsymbol{0},\omega_{\theta m}^2\mathbf{I}_{K-2})$. Hence, linear predictors in \eqref{eq:m1} can be expressed in the following way:  
	\begin{equation}\label{eq:model2}
		\boldsymbol\theta = \boldsymbol{1}_S\beta_{0\theta}+\mathbf{X}\boldsymbol{\beta}_\theta+\sum_{m=1}^M\tilde{\mathbf{B}}_{m}\boldsymbol{\tilde{\gamma}}_{\theta m}+\mathbf{u}_\theta.
	\end{equation}
	The model specification can be completed by the already discussed priors \eqref{eq:priorbeta0}, \eqref{eq:priorbeta} and \eqref{eq:prioru}, whereas for the scaling parameter $\omega_{\theta m}$ the same prior of equation \eqref{eq:prior_spl} can be set. In this way, a standard Bayesian P-splines model can be implemented, even if only proper priors are specified. 
	
	\subsubsection{Variable selection: the grouped horseshoe prior}
	When several covariates are available and their relationships with the modeled latent parameters are unknown, it can be useful to set a prior distribution that is able to shrink the non-relevant regressors to zero. \cite{scheipl2012spike} proposed to use spike-and-slab priors for functional selection. The behavior of such priors is also mimicked by the HS priors, which have been proposed in a hierarchical version to deal with shrinkage of grouped regression terms \citep{xu2016bayesian}. Since all the coefficients related to a covariate can be considered to form a group, a grouped HS prior appears suitable to be applied in this framework, rearranging the model in \eqref{eq:model2} to 
	\begin{equation}\label{eq:model3}
		\boldsymbol\theta = \boldsymbol{1}_S\beta_{0\theta}+\sum_{m=1}^M\mathbf{Z}_{m}\boldsymbol{\alpha}_{\theta m}+\mathbf{u}_\theta,
	\end{equation}
	where $\mathbf{Z}_{m} = [\mathbf{x}_{\bullet m}\ \ \tilde{\mathbf{B}}_{m}]\in \mathbb{R}^{S\times (K-1)}$ and $\boldsymbol\alpha_{\theta m}=\left(\beta_{\theta m}, \tilde{\boldsymbol{\gamma}}_{\theta m}\right)\in\mathbb{R}^{K-1}$. To implement the grouped HS prior, the following hierarchy is necessary:
	\begin{equation}
		\begin{aligned}
			&\boldsymbol{\alpha}_{\theta m}|\boldsymbol{\delta}_{\theta m},\lambda_{\theta m},\eta_\theta \sim\mathcal{N}_{K-1}(\boldsymbol{0},\eta_\theta^2\lambda_{\theta m}^2\text{diag}[\boldsymbol{\delta}_{\theta m}]),\ m=1,\dots M;\\
			&\delta_{\theta km}\sim\mathcal{C}^+(0,1),\ \ k=1,\dots,K-1;\ \  m=1,\dots M; \\
			&\lambda_{\theta m}\sim\mathcal{C}^+(0,1),\ m=1,\dots M;\\
			&\eta_\theta\sim\mathcal{C}^+(0,s_{\hat\theta});
		\end{aligned}
	\end{equation}
	where $\boldsymbol{\delta}_{\theta m} = \left(\delta_{\theta 1m},\dots,\delta_{\theta (K-1)m}\right)^\top$ and $\mathcal{C}^+(\cdot,\cdot)$ denotes an half-Cauchy distribution. The parameter $\eta_\theta$ represents the global scale of the regression coefficients. For this reason, its prior scale is set equal to the standard deviation of posterior estimates obtained from the station-specific exploratory GEV models, to account for the different magnitudes of the modeled quantities. The prior hierarchy is completed by a covariate-specific scale $\lambda_{\theta m}$ that controls the relevance of the whole effect and the coefficient-specific parameter $\delta_{\theta km}$. 

	\subsection{Posterior inference and model comparison}\label{sec:post}
	
	As previously mentioned, an MCMC approach is adopted to draw $B$ samples from the posterior distributions of the model parameters, by exploiting the \texttt{Stan} probabilistic language. After draws from the posteriors of the basic parameters are obtained, it is possible to consequently retrieve other posterior distributions of useful quantities. More in detail, it is possible to have the posterior for the generic GEV parameter related to station $s$: $\theta_s|\mathbf{y}$. If the interest is on an out-of-sample location $s^\prime$, such quantity cannot be computed due to the presence of the station-specific random effect term. To propagate the uncertainty, it is possible to obtain a prediction of the GEV parameter defined as $\tilde{\theta}_{s^\prime}|\mathbf{y}=\left(\beta_{0\theta}+f(\mathbf{x}_{s^\prime\bullet}^T)+\tilde{u}_{\theta}\right)|\mathbf{y}$. The function of the covariates $f(\mathbf{x}_{s^\prime\bullet}^T)$ depends on the kind of model that is analyzed (linear or spline regression) and the $b$-th replicate of random effect term is generated as $\tilde{u}_{\theta}^{(b)}\sim\mathcal{N}\left(0,{\kappa_\theta^2}^{(b)}\right)$, where ${\kappa_\theta^2}^{(b)}$ is the $b$-th draw from $\kappa_\theta^2|\mathbf{y}$. The samples from $\theta_s|\mathbf{y}$ or $\tilde{\theta}_{s^\prime}|\mathbf{y}$ can be combined by following the \eqref{eq:rp} to have posterior distributions of the return period denoted with $Q_{1/R,s}|\mathbf{y}$ or $\tilde Q_{1/R,s^\prime}|\mathbf{y}$ for the estimated and the predicted ones, respectively.
	
	The posterior predictive distribution is another important quantity for making predictions and model assessments. It is possible to recover a random sample from it by exploiting the MCMC posterior samples of GEV parameters. In particular, the $b$-th replicate from the posterior predictive $y_{st}^{rep}|\mathbf{y},\forall s,t,$ is obtained generating from:
	$
	{y_{st}^{rep}}^{(b)}\sim GEV(\mu_s^{(b)},\sigma_s^{(b)},\xi_s^{(b)})$. Similarly, the posterior predictive distributions for out-of-sample stations, denoted as $\tilde y_{s^\prime t}^{rep}|\mathbf{y},\forall s^\prime,t$, can be drawn relying on the posteriors $\tilde{\theta}_{s^\prime}|\mathbf{y}$.
	
	The posterior predictive distribution constitutes the pillar of several model performance evaluation tools that are listed hereafter.  In particular, as shown in the next section, a cross-validation study is carried out to assess and compare the performances of the models. The quantities that are introduced in the following are computed by relying on the posterior predictive $\tilde y_{s t}^{rep}|\mathbf{y}_{-s}$, i.e., obtained after fitting a model without observations from station $s$.
	
	To evaluate the calibration of predictions produced by Bayesian models, the probability integral transforms (PIT) are widely used \citep{dawid1984present}. In particular, they are defined as
	\begin{equation}
		PIT_{st}=\mathbb{P}\left[\tilde y_{s t}^{rep}<y_{st}|\mathbf{y}_{-s}\right],
	\end{equation}
	i.e. the cumulative probability of the posterior predictive distribution up to the observed value $y_{st}$. If the model predictions are calibrated, PIT values follow a uniform distribution.
	Bayesian p-values constitute another useful posterior predictive check. They can be flexibly defined, depending on the inferential goal characterizing the procedure. In extreme value estimation, GEV quantiles represent an important target quantity, since they determine return levels. For this reason, station-specific Bayesian p-values are defined for a given return period $R$:
	\begin{equation}
		\text{P-val}_{R,s} = \mathbb{P}\left[\tilde Q_{1/R,s}<q_{1/R}(\mathbf{y}_s)|\mathbf{y}_{-s}\right],
	\end{equation}
	where $q_{1/R}(\mathbf{y}_s)$ is the sample quantile of the maxima of station $s$. In this case, good model performances are underlined by values of $\text{P-val}_{s}$ nearby 0.5.
	
	Lastly, the continuous ranked probability score (CRPS) is largely used to evaluate the probabilistic predictions under continuous densities, even in the extreme values literature 
	\citep{friederichs2012forecast}. It is a score computed specifically for each observation $y_{st}$ and it is indicated with $\text{CRPS}(y_{st})$, and the \texttt{R} package \texttt{scoringRules} can be exploited to evaluate it \citep{crps}. Note that the model showing lower scores is preferable in terms of calibration and sharpness of the predictions.

	\begin{figure}
		\centering
		\includegraphics[width = \linewidth]{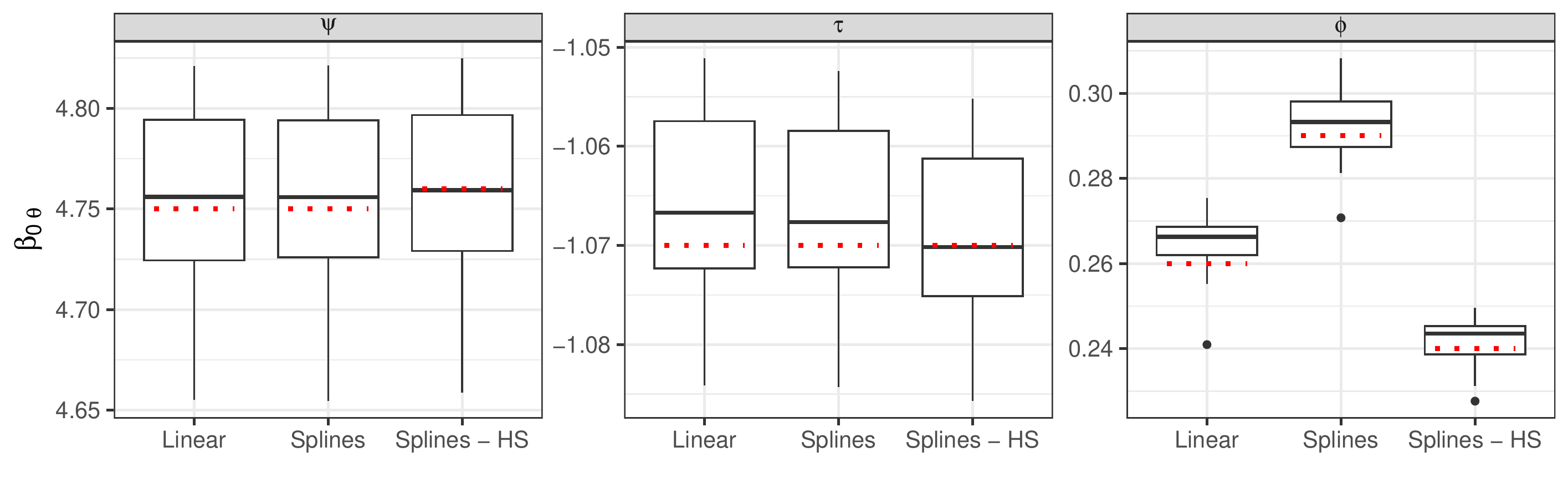}
		\caption{Boxplot of the $G=31$ posterior means from the cross-validation compared to the estimate obtained in the model fitted with all the stations. }
		\label{fig:stab_beta}
	\end{figure}
	
	\section{Application}\label{sec:empirical}
	
	The modeling strategies described in Section~\ref{sec:model} are applied to the Danube basin data introduced in Section~\ref{sec:data}. In particular, results about three different Bayesian models are compared: the one assuming linear effects for the covariates, labeled as \textit{Linear} and defined by equations in \eqref{eq:lin}, the basic P-spline model of \eqref{eq:m1}, labeled as \textit{Splines}, and its extension to automatically perform model selection through grouped horseshoe priors (labeled as \textit{Splines-HS}). 
	
	To assess the performances of the considered models, results from a folded cross-validation study are reported in Section~\ref{sec:CV}, whereas the outcomes from the analysis carried out on the full dataset are discussed in Section~\ref{sec:results}.

	\subsection{Cross-validation study}\label{sec:CV}
	
	The whole set of $S=62$ stations is randomly partitioned into $G=31$ groups constituted by a couple of stations each. A folded cross-validation study is executed, by repeatedly fitting the 3 compared models and excluding a couple of stations at each iteration. The quantities introduced in Section~\ref{sec:post}, PIT and CRPS in particular, are evaluated on the out-of-sample stations.
	
	A first indication from the folded cross-validation study concerns the stability of the estimates with respect to the removal of stations. Given that the models are characterized by different parameterizations, the intercepts $\beta_{0\theta}$ are taken into consideration for this aspect. Figure~\ref{fig:stab_beta} compares the distributions of the posterior means obtained in the 31 runs to the estimates of the intercepts in the models fitted considering all the stations. The estimation of such parameters seems to be stable: the estimates obtained exploiting the full dataset are often close to the median of the distribution and, in general, included in the boxes. The only exception concerns the $\boldsymbol\phi$ parameter under \textit{Linear}, remaking that such parameter is also characterized by evident differences in the estimates across the models. This could be expected due to the difficulties in identifying the shape parameter.

	\begin{figure}
		\centering
		\includegraphics[width = \linewidth]{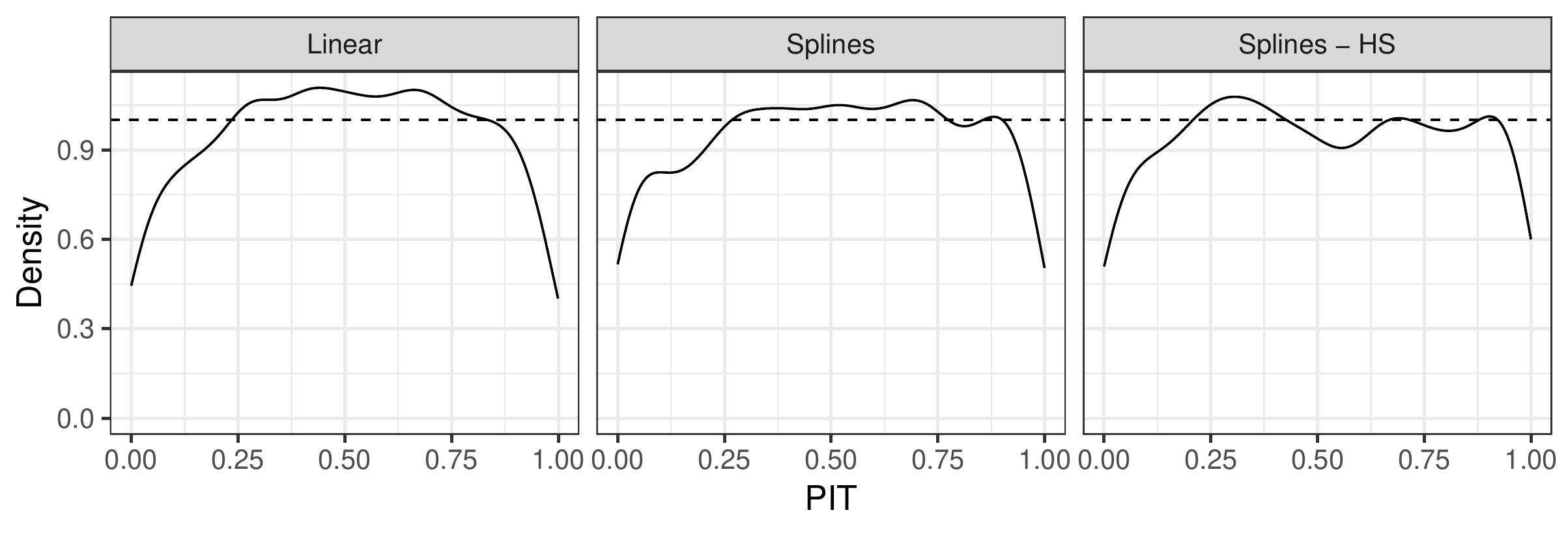}
		\caption{Kernel densities of the distribution of $PIT_{st},\ \forall s,t$, density of the Uniform distribution as dashed line.}
		\label{fig:PIT}
	\end{figure}

	The calibration of the predictions produced by the compared models is firstly evaluated by exploring the distribution of $PIT_{st}$, recalling that a uniform distribution is required for a calibrated model. The kernel densities are shown in Figure~\ref{fig:PIT}, compared with the expected uniform distribution. In the models including flexible regression terms (\textit{Splines} and \textit{Splines-HS}), PIT distributions are more compliant with the Uniform if compared to the \textit{Linear} model, where the excess of values far from 0 or 1 is more evident.
	Another indication about the calibration of predictions can be deduced from the Bayesian p-values $\text{P-val}_{R,s}$. To summarise them, $\text{P-val}_{R}^*$ denotes the proportion of Bayesian p-values far from the extremes, i.e. included in the interval $(0.05,0.95)$. Selecting $R=50$, $\text{P-val}_{R}^*$ is equal to 0.85 for \textit{Splines}, 0.87 for \textit{Splines-HS} and 0.90 for \textit{Linear}; concerning $R=100$, a value of 0.87 is observed for \textit{Splines}, and 0.92 for \textit{Linear} and \textit{Splines-HS}. To sum up, the three models show good results in terms of calibration of predictions, with the \textit{Linear} slightly penalized in terms of PIT and the \textit{Splines} model in terms of Bayesian p-values for quantiles predictions. 
	
	\begin{figure}
		\centering
		\includegraphics[width = \linewidth]{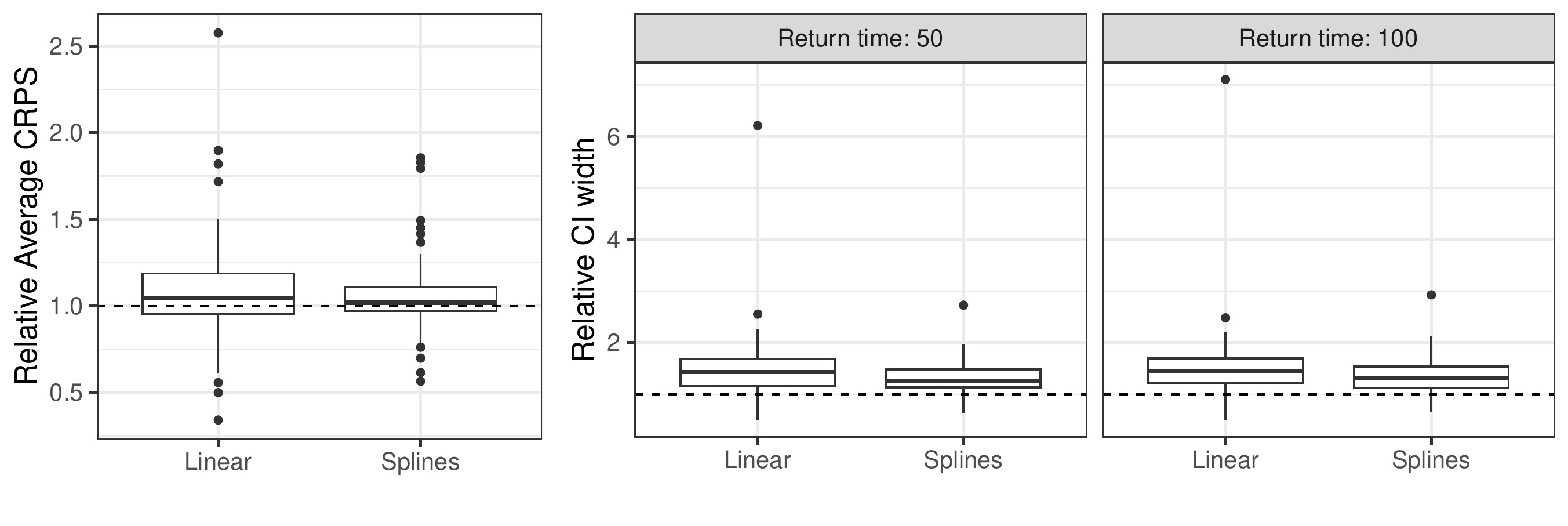}
		\caption{Left-hand-side plot: box-plot of ACRPS$_s$. Right-hand-side: box-plot of the widths of 90\% credible intervals for $\tilde{Q}_{1/R,s}|\mathbf{y}_{-s}$. In both cases, the values are divided by the corresponding ones obtained under the \textit{Splines-HS} model (benchmark).}
		\label{fig:ccrps}
	\end{figure}
	
	Lastly, further evaluations of the reliability and the sharpness of predictions are discussed. The average CRPS (ACRPS) is computed to have a station-specific summary: $\text{ACRPS}_s=T^{-1}\sum_t\text{CRPS}(y_{st})$, and their distributions across the stations are depicted in Figure~\ref{fig:ccrps}. To set the proposed \textit{Splines-HS} model as a benchmark, the values are relativized by dividing them for the corresponding ACRPS observed under this model. The median of the distributions of relative ACRPS is above 1 both for \textit{Linear} and \textit{Splines} models, where, respectively, 60\% and 61\% of stations have higher ACRPS than under \textit{Splines-HS} model. Note that the $61\%$ of stations have higher ACRPS under \textit{Linear} model if compared to the \textit{Splines} one, pointing out the merits of introducing flexible effects in the model. Another indication about the sharpness of prediction can be deduced from the width of the 90\% credible intervals of the posterior of quantiles $\tilde{Q}_{1/R,s}|\mathbf{y}_{-s}$, for $R=\{50,100\}$. Also in this case, Figure~\ref{fig:ccrps} reports the distribution of the station-specific widths divided by those obtained under the \textit{Splines-HS} model. It is interesting to stress how the \textit{Splines-HS} model has intervals in median the 26.6\% and 42.6\% narrower than the intervals retrieved with \textit{Splines} and \textit{Linear} models, respectively. 
	
	\begin{table}[]
		\centering
		\begin{tabular}{@{}lllllll@{}}
			\toprule
			& \multicolumn{2}{c}{$\kappa_\psi$} & \multicolumn{2}{c}{$\kappa_\tau$} & \multicolumn{2}{c}{$\kappa_\phi$} \\ \cmidrule(l){2-3}  \cmidrule(l){4-5}  \cmidrule(l){6-7} 
			& Mean      & 95\% C.I.         & Mean      & 95\% C.I.         & Mean      & 95\% C.I.         \\ \midrule
			Linear       & 0.47   & {[}0.39,0.58{]}   & 0.15    & {[}0.09,0.21{]}   & 0.05    & {[}0.00,0.12{]}   \\
			Splines      & 0.34    & {[}0.26;0.44{]}   & 0.03    & {[}0.00,0.09{]}   & 0.03    & {[}0.00,0.10{]}   \\
			Splines - HS & 0.27   & {[}0.16,0.42{]}   & 0.04    & {[}0.00,0.10{]}   & 0.04    & {[}0.00,0.11{]}   \\ \bottomrule
		\end{tabular}
		\caption{Posterior estimates related to the random effects scale parameters.}
		\label{tab:reff}
	\end{table}
	
	\subsection{Results}\label{sec:results}

	According to the results of the folded cross-validation study presented in the previous section, allowing for non-linear relationships among covariates and GEV parameters leads to some gains in terms of predictive ability. These improvements are even more noticeable when a prior able to automatically execute the variables selection step is assumed. Similar conclusions can also be detected by comparing the models fitted relying on the whole dataset. As an overall measure of goodness of fit, the leave-one-out information criterion \citep[LOOIC,][]{vehtari2017practical}  is considered and it registers the lowest values for the \textit{Splines-HS} model (22219.5), followed by the \textit{Splines} one (22222.1) and, more detached, the \textit{Linear} model (22245.7).
	
	\begin{figure}
		\centering
		\includegraphics[width = \linewidth]{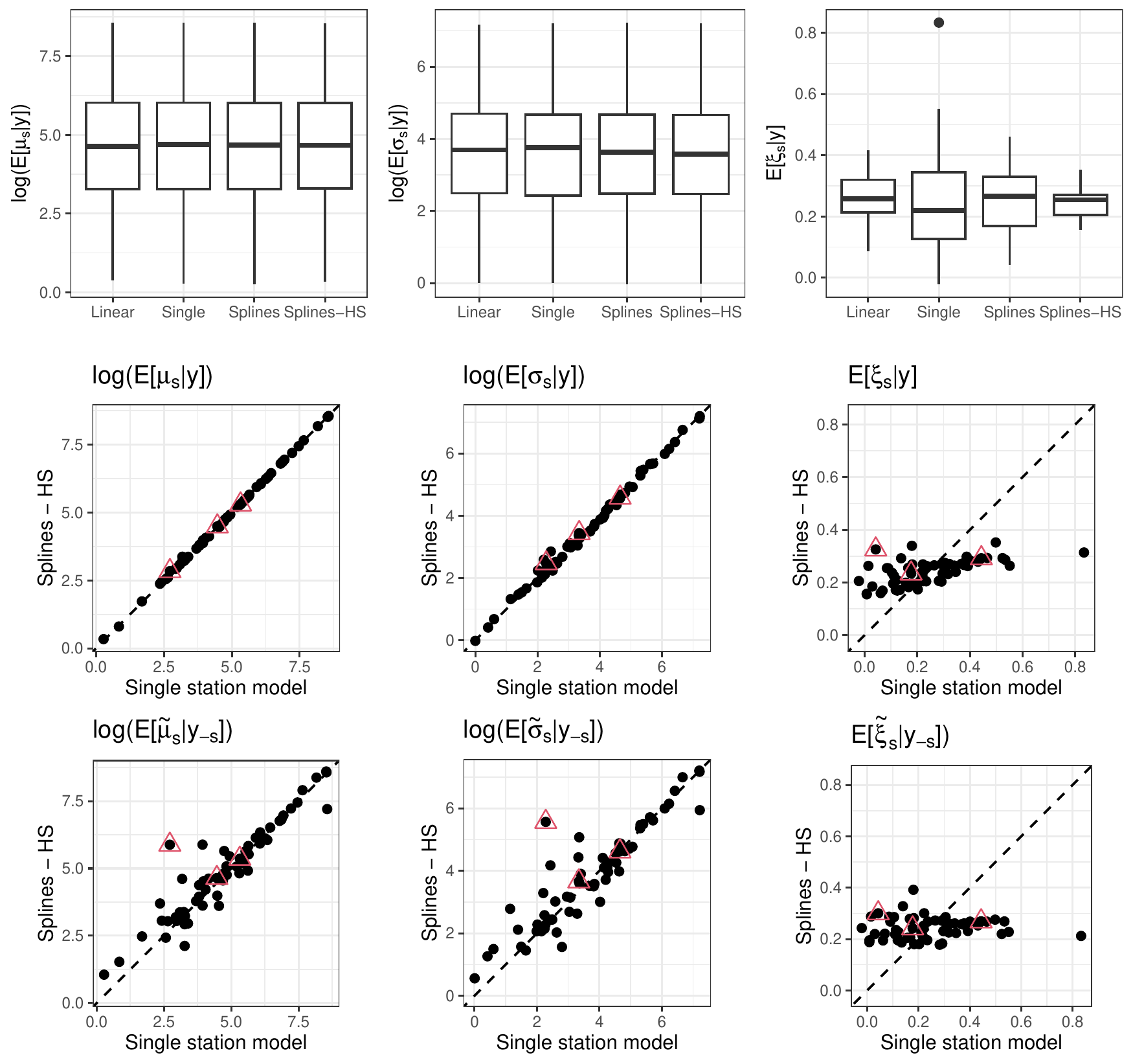}
		\caption{Boxplots of GEV parameters posterior means under the considered models (first row). Comparison between estimates (second row) and out-of-sample predictions (third row) from the \textit{Splines-HS} model and the station-specific ones. The red triangles indicate the stations considered for return levels of Figure~\ref{fig:rl}. }
		\label{fig:comparison}
	\end{figure}
	
	A first insight to understand the benefits led by the models with Bayesian P-splines can be deduced from Table~\ref{tab:reff}, reporting the posterior summaries about the random effects scale parameters $\kappa_{\theta}$. Such quantities can be considered as measures of the amount of signal captured by the covariates in the regression models: the higher the values, the lower the variability explained by the covariates. The \textit{Linear} model registers noticeably higher scales, especially for the random effects related to parameters $\boldsymbol{\psi}$ and $\boldsymbol{\tau}$. Despite such differences, it is interesting to remark that the in-sample estimates of the stations-specific GEV parameters $\mu_s$ and $\sigma_s$ are similar across the considered models, whereas differences can be observed for the shape parameter $\xi_s$, for which the models induce different levels of shrinkage. These results are depicted by the boxplots in the first row of Figure~\ref{fig:comparison}, where also the estimates obtained under the station-specific models are added for benchmarking purposes. As a consequence, the inflation of the scales $\kappa_\theta$ might lead to over-dispersed out-of-sample predictions: such behavior is captured by the PIT distribution previously reported in Figure~\ref{fig:PIT} and the general increase of the width of the credible intervals (Figure~\ref{fig:ccrps}).

\begin{figure}
		\centering
		\includegraphics[width = \linewidth]{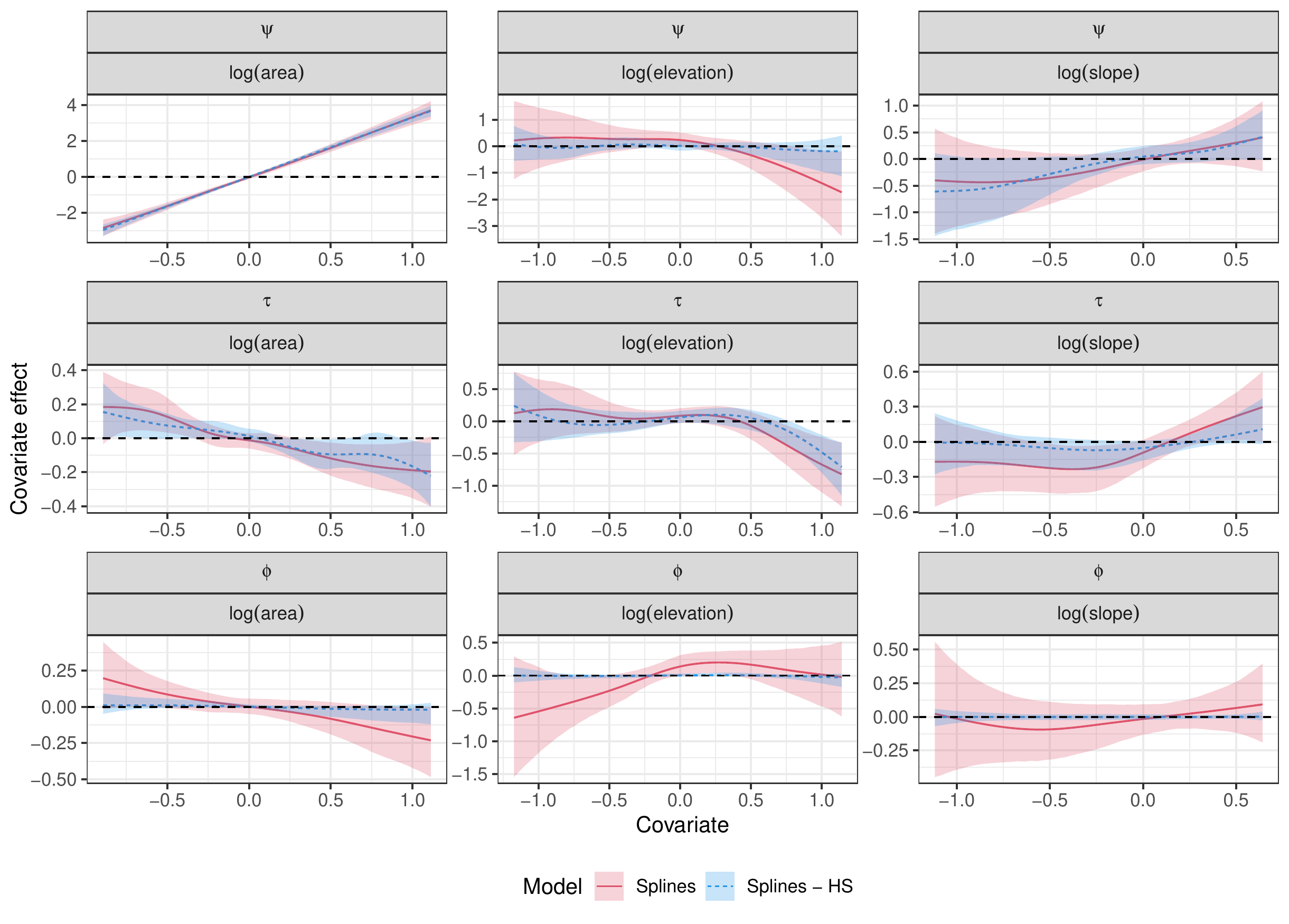}
		\caption{Covariate effects estimated for three selected covariates under models \textit{Splines} and \textit{Splines-HS} for the three functionals of GEV parameters. Shaded areas depict the 90\% credible intervals.}
		\label{fig:effects}
	\end{figure}

	Let now shift the focus to the comparison between the two models that include flexible regression terms. Figure~\ref{fig:effects} shows how three selected covariates (area, elevation and slope) impact on the transformations of GEV parameters. As a first cue, it can be noticed the marked non-linearity of several effects (elevation and slope, primarily). The trends detected by the two models are similar, however, the impact of the grouped HS prior for the splines coefficients emerges. The shrinkage towards zero for negligible effects is evident under the \textit{Splines-HS} model, especially when modeling $\boldsymbol\phi$, i.e. the function of the shape parameter. In this case, the \textit{Splines} model individuates trends endowed with considerably higher uncertainty, producing intervals that include the 0 value almost everywhere. The decrease in the effect uncertainty is also detectable when modeling parameters $\boldsymbol\psi$ and $\boldsymbol\tau$, even if less pronouncedly. Observing the effect of area on the location parameter $\boldsymbol{\psi}$ it can be pointed out that the grouped HS prior is also able to firmly lead back the flexible effect to the linearity assumption.
	
	The results concerning covariate effects can be put in relationship with those about the random effects scales shown in Table~\ref{tab:reff}. Indeed, the combination of these outputs allows to motivate the lower dispersion of the station-specific estimates of the shape parameters $\xi_s$ under the \textit{Splines-HS} model, already noticed in boxplots of the first row in Figure~\ref{fig:comparison}. On the other hand, the \textit{Splines} model produces scattered estimates. In light of the massive shrinkage induced by the grouped HS priors, such variability of estimates could be poorly supported by the data, possibly leading to problems of instability. In fact, it is widely known that the identification of the shape parameter of the GEV distribution is a tricky task \citep[see, e.g.,][]{johannesson2022approximate}, and the grouped HS prior can help in avoiding over-fitting in this framework.
	
	The second and the third rows of Figure~\ref{fig:comparison} allow us to deepen the connections between parameters estimates obtained with the \textit{Splines-HS} model and the GEV distribution fitted on the single stations. The results related to the in-sample estimates confirm that no relevant differences are detected in estimating $\mu_s$ and $\sigma_s$, they provide further evidence about the strong shrinking process affecting the estimates of $\xi_s$, which are gathered around 0.25. It is also interesting to explore how the GEV parameters are predicted when data related to the station are excluded from the fitting sample, taking the outcome of the folded cross-validation study (third row). As expected, the predictions concerning $\mu_s$ and $\sigma_s$ are more scattered with respect to the estimates  from the single-station models, even if the correlation between estimates and predictions is strong. From these diagnostic plots, three stations, whose points are embedded in a red triangle, are selected to investigate the inference on return levels through the different modeling strategies. To this aim, the distances between predictions and single-station model estimates are considered, taking the stations having maximum (\#6242530), median (\#6243240) and minimum (\#6342610) distances, noting that such stations are also representative of different values of the shape parameter according to the single-station models. 
	
	\begin{figure}
		\centering
		\includegraphics[width = \linewidth]{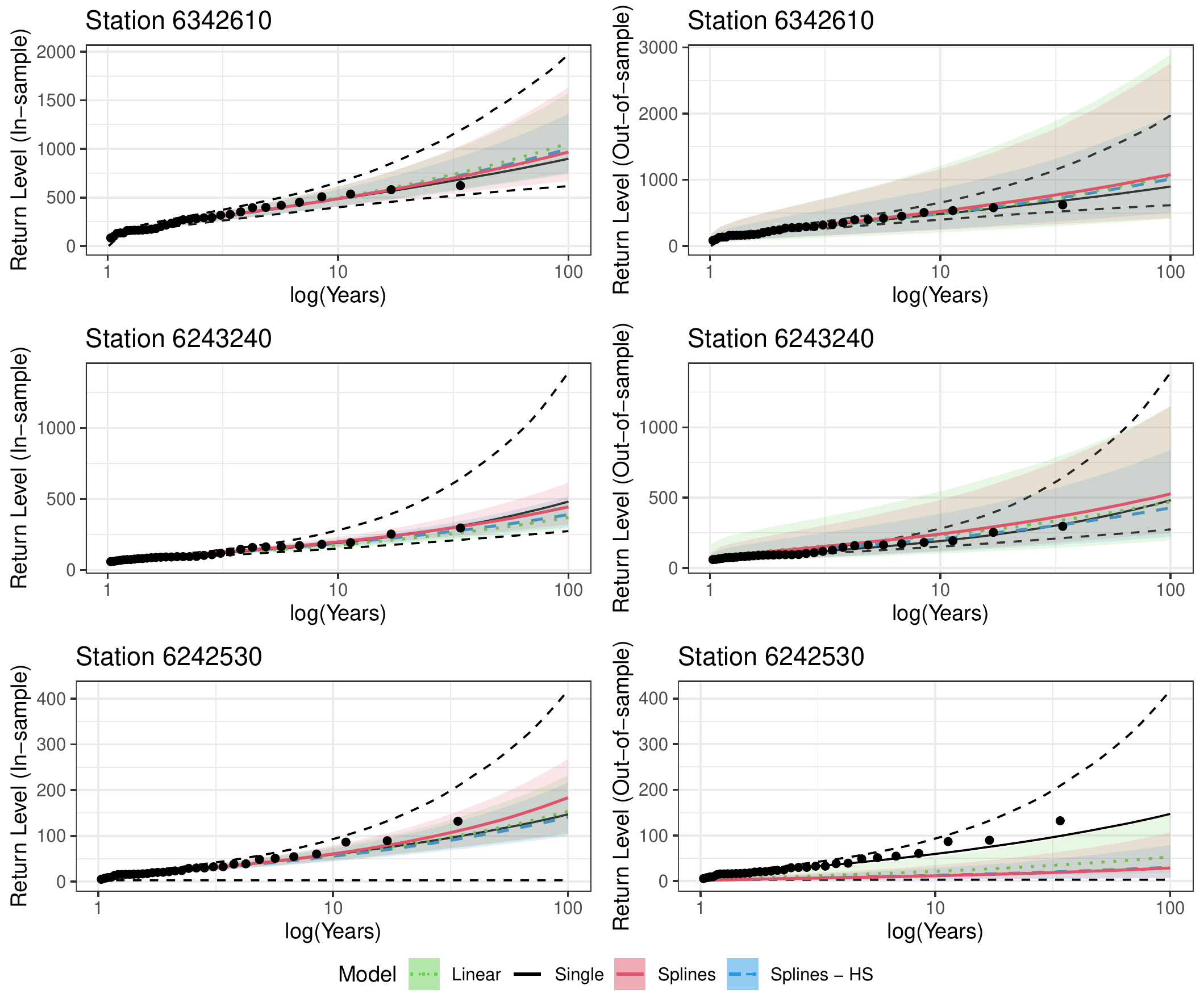}
		\caption{Return levels of river discharge (in $m^3/s$) estimated with the whole sample (left column) and excluding the observation available from the station (right column). The shaded areas indicate the 90\% credible interval. The points represent the ordered observations.}
		\label{fig:rl}
	\end{figure}

	To complete the analysis of the results, a brief discussion on the estimates and the out-of-sample predictions of river discharge return levels is carried out (outcomes reported in Figure~\ref{fig:rl}). As expected, the in-sample estimates are generally characterized by lower levels of uncertainty than predictions, whose variability is inflated by the presence of random effects generated from the prior, as described in Section~\ref{sec:post}. Another general trend to point out is that the single-station models produce estimates with larger credible intervals, mainly due to the issues in estimating the shape parameters. Conversely, the models fitted on the overall basin allow borrowing strength across the stations, reducing such variability through the aforementioned shrinkage process on $\xi_s$. Besides, as already pointed out in Section~\ref{sec:CV}, the \textit{Splines-HS} model is also able to produce return level estimates with lower uncertainty levels than the other strategies, by combining lower variability in effects identification (Figure~\ref{fig:effects}) and lower random effects scale parameters (Table~\ref{tab:reff}). Despite the narrower bands, the points representing the observed values are included in the credible intervals, with the exception of predictions for station \#6242530, i.e. the one characterized by the maximum distance between predicted and estimated parameters.

	\section{Concluding remarks}\label{sec:concl}
	
	This paper aims at illustrating the potential of Bayesian models in introducing flexibility in extreme value analysis. In particular, the linearity assumption, often restrictive in dealing with complex phenomena such as environmental ones, is relaxed proposing non-linear functional relationships. Furthermore, a suitable regularizing prior is introduced, allowing the incorporation of variables and functional selection steps within the model. The use of the popular \texttt{Stan} software to sample from the posteriors could also foster practitioners in using more sophisticated statistical techniques. 
	
	The performances of the models considered in the paper are compared by means of a cross-validation study that evaluates their ability in predicting return levels at ungauged locations. In doing so, the advantages brought by the use of splines regression tied with a regularizing prior can be highlighted. Indeed, its use allows us to sensibly reduce the uncertainty of the predictions without affecting model calibration if compared to other considered model specifications. 
	
	Despite the application tackles extreme value analysis from the block-maxima perspective, by adopting the typical GEV distribution, the underlying idea of setting a semi-parametric regression with regularizing priors can also be extended to other distributional assumptions and approaches of extreme value theory. Among the others, we mention the Blended-GEV by \cite{castro2022practical}, which solves the GEV problem of having a finite lower tail when the shape parameter is positive, or the widespread peak-over-threshold approach. In the latter framework, the proposed strategy might help in both the threshold determination step and in the analysis of the exceedances through the Generalized Pareto distribution. 
	
	Lastly, it is worth stressing that the principle behind the use of a prior encouraging a grouped variable selection can also be extended to other low-rank structure matrices such as tensors, useful to model a spatially structured effect, interactions and also categorical variables \citep{scheipl2012spike}.

	\section*{Acknowledgments}
	
	The work of Aldo Gardini was partially supported by MUR on funds FSE REACT EU - PON R\&I 2014-2020 and PNR (D.M. 737/2021) for the RTDA\_GREEN project (title: "Modelli statistici per lo studio della convergenza spaziale verso la transizione verde", J41B21012140007).
	\bibliographystyle{abbrvnat}
	\bibliography{bibliography}
	
\end{document}